\documentclass{article}

\usepackage{mathrsfs}
\usepackage{amsfonts}
\usepackage{amsmath}
\usepackage{amssymb}
\usepackage{graphicx}

\title{Joining inner space to outer space}
\author{Gordon McCabe}

\begin{document}

\maketitle

\begin{abstract}

The purpose of this paper is to demonstrate that it is possible, in
principle, to obtain knowledge of the entire universe at the present
time, even if the radius of the universe is much larger than the
radius of the observable universe.

\end{abstract}

\section{The construction}

Cosmologist George Ellis states that ``Our view of the universe is
limited by the visual horizon, comprised of the worldlines of [the]
furthest matter we can observe - namely, the matter that emitted the
CBR [Cosmic Background Radiation] at the time of last scattering\dots
Visual horizons do indeed exist, unless we live in a small universe,
spatially closed with the closure scale so small that we can have
seen right around the universe since decoupling," (Ellis 2007,
section 2.4.2). He asserts that ``Claims about what conditions are
like on very large scales - that is, much bigger than the Hubble
scale - are unverifiable, for we have no observational evidence as to
what conditions are like far beyond the visual horizon," (ibid.,
section 4.3). The purpose of this paper is to devise a model of our
universe in which these assertions would be false.

Let us begin by making the following assumptions:

\begin{enumerate}
\item{The Friedmann-Robertson-Walker models of general relativistic cosmology are correct,
but the spatial universe in these models is a 3-dimensional compact
manifold-with-boundary $\Sigma$. Compactness entails that the spatial
universe is of finite volume, (see Cornish and Weeks, 1998).}
\item{The radius of the spatial universe at the current time is many
orders of magnitude greater than the radius of the observable spatial
universe.}
\item{In particle physics, string theory is false, but the preon hypothesis is correct
in the sense that there is only one type of ultimate elementary
particle. Hence, all the quarks and leptons, all the gauge bosons,
and all the Higgs bosons (if they exist), are ultimately composed of
preons (see Bilson-Thompson \textit{et al}, 2006), and the standard
model emerges from the theory of preons.}
\end{enumerate}

The model universe I wish to propose is one in which outer space is
joined to inner space, in the sense that each elementary particle
contains the universe to which the elementary particle itself
belongs. This is perfectly well-defined topologically.

Let us suppose that each preon has a Compton wavelength\footnote{The
Compton wavelength of a particle of mass $m$ is $h/mc$, where $h$ is
Planck's constant, and $c$ is the speed of light.} of $r$. The
Compton wavelength can be thought of as the length scale at which
relativistic quantum field theory becomes relevant for a particle.
Hence, as a first approximation which neglects quantum field theory,
let us represent each preon as a solid ball $\mathbb{D}_i(r)$ of
radius $r$, embedded in 3-dimensional space $\Sigma$. Given the
assumption that the spatial universe is of finite volume, there will
be a finite number $N$ of preons in the universe, so they can be
enumerated $i=1,....,N$. Under the approximation made here, the
boundary of each preon is homeomorphic to the 2-sphere, $\partial
\mathbb{D}_i \cong S^2, \; i=1,...,N $. Let us also suppose that the
boundary of large-scale 3-dimensional space is homeomorphic to the
boundary of each preon, which in this case is the 2-sphere, so that
$\partial \Sigma \cong S^2$.

Now excise from $\Sigma$ the interior of each solid ball
$\mathbb{D}_i$. This means taking the complement $\Sigma_\flat =
\Sigma - \bigcup_i \text{Int}(\mathbb{D}_i)$. So doing obtains a
3-manifold-with-boundary $\Sigma_\flat$ containing numerous `holes'.

Now identify the original boundary $\partial \Sigma$ with the
boundary of each hole $\partial \mathbb{D}_i$, to obtain a quotient
topological space $\Sigma_\sharp$. In other words, define $N$
bijective identification maps $\phi_i:\partial \Sigma \rightarrow
\partial \mathbb{D}_i$, and introduce an equivalence relationship $\mathscr{R}$
which treats $p \in \partial \Sigma$ as equivalent to each $\phi_i(p)
\in \partial \mathbb{D}_i$ for $i = 1,...,N$. Points not belonging to
any component of the boundary of $\Sigma_\flat$ are each treated as
an equivalence class containing a single member. The quotient space
is the set of all equivalence classes, $\Sigma_\sharp =
\Sigma_\flat/\mathscr{R}$. There is a projection mapping from
$\Sigma_\flat$ onto $\Sigma_\sharp$, and the quotient topology is
defined to be the largest topology one can bestow on $\Sigma_\sharp$
which still permits the projection mapping to be continuous.

In such a space-time, each elementary particle is simply an embedding
of the universe within itself. If one tries to probe the inside of an
elementary particle, then one is probing inside the entire universe,
for the boundary of the elementary particle (preon) is also the
outer-most boundary of large-scale space-time. In such a universe,
one really can see the universe in a grain of sand.

If such were the case, then we would effectively have both an
internal perspective on the universe, extending out to the
conventional visual horizon, and we would potentially have an
external perspective, which would allow us to see regions of space
beyond our conventional visual horizon. Our observations would still
be restricted to the past light cone, but if the past light cone
included an elementary particle, as defined here, then the
information stored within such an elementary particle would include
everything there is to know about the universe, and would therefore
provide us, in principle, with knowledge of those parts of the
universe which lie beyond the radius of the observable universe, as
conventionally defined.

\end{document}